\documentclass[12pt,preprint]{aastex}

\slugcomment{to be sumbitted to ApJ}

\shorttitle{A HST/ACS Search for Brown Dwarf Binaries in the Pleiades}
\shortauthors{Bouy et al.}

\begin{document}


\title{A Hubble Space Telescope ACS Search for Brown Dwarf Binaries in the Pleiades Open Cluster}


\author{H. Bouy}
\affil{Instituto de Astrof\'\i sica de Canarias, C/ V\'\i a L\'actea, s/n, E-38200 - La Laguna, Tenerife, Spain}
\email{bouy@iac.es}

\author{E. Moraux}
\affil{Institute of Astronomy, University of Cambridge, Madingley Road, Cambridge. CB3 0HA, United Kingdom}
\email{moraux@ast.cam.ac.uk}

\author{J. Bouvier}
\affil{Laboratoire d'Astrophysique de l'Observatoire de Grenoble, 414 rue de la piscine, F-38400 Saint Martin d'H\`ere, France}
\email{Jerome.Bouvier@obs.ujf-grenoble.fr}

\author{W. Brandner}
\affil{Max-Planck Institut f\"ur Astronomie, K\"onigstuhl 17, D-69117 Heidelberg, Germany}
\email{brandner@mpia.de}

\author{E.~L. Mart\'\i n }
\affil{Instituto de Astrof\'\i sica de Canarias, C/ V\'\i a L\'actea, s/n, E-38200 - La Laguna, Tenerife, Spain}
\affil{University of Central Florida, Department of Physics, PO Box 162385, Orlando, FL 32816-2385, USA}
\email{ege@iac.es}

\author{F. Allard}
\affil{Centre de Recherche Astronomique de Lyon (UML 5574), Ecole Normale Sup\'erieure, 69364 Lyon Cedex 07, France}
\email{fallard@ens-lyon.fr}

\author{I. Baraffe}
\affil{Centre de Recherche Astronomique de Lyon (UML 5574), Ecole Normale Sup\'erieure, 69364 Lyon Cedex 07, France}
\email{ibaraffe@ens-lyon.fr}

\author{M. Fern\'andez}
\affil{Max-Planck Institut f\"ur Astronomie, K\"onigstuhl 17, D-69117 Heidelberg, Germany}
\email{matilde@mpia.de}


\clearpage
\begin{abstract}
We present the results of a high-resolution imaging survey for brown dwarf binaries in the Pleiades open cluster. The observations were carried out with the Advance Camera for Surveys \citep{ACS_INSTR_HANDBOOK} onboard the Hubble Space Telescope. Our sample consists of 15 bona-fide brown dwarfs. We confirm 2 binaries and detect their orbital motion, but we did not resolve any new binary candidates in the separation range between 5.4~AU and 1700~AU and masses in the range 0.035--0.065~M$_{\sun}$. Together with the results of our previous study \citep{2003ApJ...594..525M}, we can derive a visual binary frequency of 13.3$^{+13.7}_{-4.3}$\% for separations greater than 7~AU masses between 0.055--0.065~M$_{\sun}$ and mass ratios between 0.45--0.9$<q<$1.0. The other observed properties of Pleiades brown dwarf binaries (distributions of separation and mass ratio) appear to be similar to their older counterparts in the field. 
\end{abstract}


\section{Introduction}
Young open clusters offer the advantage that both the age and distance are precisely known so that brown dwarfs candidates are more easily identified from their positions in colour-magnitude diagrams, relative to the expected position of the cluster's sub-stellar isochrone. Over the last few years, a large number of authors have published results of large surveys looking for substellar members of the Pleiades \citep{2003MNRAS.343.1263N,2003A&A...400..891M,2002MNRAS.335..687D,2002MNRAS.335..853J}. Using theoretical models \citep{2000ApJ...542..464C}, the magnitude of an object can be readily converted to a mass (given the age and distance of the cluster) and the resulting IMF estimated. While detailed studies of the IMF of the Pleiades' very low mass stars and brown dwarfs have already been performed \citep[see e.g][]{2002MNRAS.335L..79D, 2002MNRAS.335..853J,1999MNRAS.303..835H}, the contribution of multiple systems to the IMF has rarely been taken into account. In this study, we obtained high angular resolution images of a sample of brown dwarfs in the Pleiades cluster in order to investigate the occurrence of multiple systems among sub-stellar objects, and its implications on:
\begin{enumerate}
\item the formation and evolution processes of brown dwarfs
\item the properties of these multiple systems in comparison with those of the field and in star forming regions. 
\item the contribution of sub-stellar objects to the IMF
\end{enumerate}

The Pleiades is one of the best studied open clusters. Its age \citep[105-140~Myr][]{martin_cancun2006} and distance \citep[d=135~pc, see e.g][]{2004Natur.427..326P, 2004A&A...418L..31M} are well known and its IMF has been well studied over the stellar mass range. All the targets come from the same star forming region: they formed under similar initial conditions and are now following identical evolutionary paths, which is not the case of field brown dwarfs for which in general we know neither the age nor the distance precisely. Moreover, the Pleiades cluster offers two important advantages for our study in comparison with other clusters, star forming regions or associations. First of all, there exists a relatively large sample of confirmed brown dwarfs, which is of prime importance in making a good statistical study. Secondly because the cluster is not so far away as to exclude a search for close visual binaries. These considerations  make this cluster the ideal place for a complementary study to the field ultracool dwarfs presented by \citet{2005ApJ...621.1023S,2003AJ....126.1526B,2003ApJ...586..512B,2003ApJ...587..407C,2003AJ....125.3302G}.
 
In a first attempt to investigate brown dwarf binaries, \citet{1998ApJ...509L.113M, 2000ApJ...543..299M} surveyed 34 very low mass Pleiades members with HST and adaptive optics at CFHT. They found only one binary at a resolution of 0\farcs2 or larger (27~AU, but it failed the lithium test and was therefore not confirmed as a Pleiades member. More recently, \citet{2003ApJ...594..525M} used the HST/WFPC2 and found only four binary candidates at a resolution of $\sim$0\farcs060 or larger (8.1~AU at 135~pc) among a total sample of 25 objects. In this paper, we present the result of our complementary. higher resolution ACS observations. In section \ref{obs_strat}, we present the new sample, the observations and the data analysis. In section \ref{mutliple_syst}, we present the results on the resolved multiple systems. In section \ref{conf_photom}, we discuss the confirmed and unresolved photometric binary candidates. In sections \ref{bin_frequ} and \ref{discussion}, we calculate and discuss the binary frequency.

\section{Observational strategy and techniques \label{obs_strat}}
In order to refine the previous studies of Pleiades brown dwarfs binaries \citep{2000ApJ...543..299M, 2003ApJ...594..525M}, we used the higher angular resolution provided by HST/ACS-HRC (program SNAP-9831, P.I. Bouy). Using PSF fitting, the observations we obtained with HST/ACS allow us to resolve multiple systems with separations as low as $\sim$0\farcs040 ($\sim$5.4~AU at the distance of the Pleiades). This is more than 5 times better than the NICMOS study of \citet{2000ApJ...543..299M} and 1.5 times as good as the WFPC2/PC study of \citet{2003ApJ...594..525M}. Moreover, the sensitivity of HST/ACS in the chosen filter is $\sim$5 times greater than the WFPC2/PC \citep[see ][]{WFPC..Instrument..Handbook}. This allows us to investigate systems with close companions and with low flux ratios between the companion and the primary.\\

  \subsection{Sample}
The initial sample consists of 32 brown dwarfs (spectral types later than M7) in the magnitude range I=18.0~mag to I=22.9~mag, identified from deep, wide-field surveys of the Pleiades cluster \citep{2003A&A...400..891M, 2001A&A...367..211M, 1999MNRAS.303..835H, 1998A&A...336..490B}. Six objects (the binaries CFHT-PL-12, IPMBD 25 and IPMBD 29, and the unresolved objects CFHT-PL-15, CFHT-Pl-21 and CFHT-Pl-24) had already been observed with WFPC2 by \citet{2003ApJ...594..525M}, and two more (CFHT-Pl-11 and CFHT-Pl-13) with NICMOS by \citet{2000ApJ...543..299M}. All targets have been identified as brown dwarfs using near-infrared and optical photometry analysis and/or spectroscopy. The sample covers a mass range from 0.025 to 0.080~M$_{\sun}$ (see Table \ref{pleiades_acs_sample}). The membership of our targets has been already confirmed by proper motion measurements or spectroscopy \citep{2001A&A...367..211M, 2003A&A...400..891M}.

\begin{deluxetable}{llllllllll}
\tablecaption{Pleiades sample \label{pleiades_acs_sample}}
\tabletypesize{\scriptsize}
\tablewidth{0pt}
\tablehead{
\colhead{Name} & \colhead{R.A (2000)} & \colhead{Dec. (2000)} & \colhead{$I$ [mag]} & \colhead{$I-Z$ [mag]}
}
\startdata
{\bf \object{Cl* Melotte 22 CFHT-Pl 11}}       &       03 47 39.0      &       +24 36 22.1  	& 17.91 & \nodata \\
{\bf \object{Cl* Melotte 22 CFHT-Pl 12}$^{\star}$}       &       03 53 55.1      &       +23 23 36.4  & 17.87	& 1.04	\\
{\bf \object{Cl* Melotte 22 CFHT-Pl 13}}       &       03 52 06.72     &       +24 16 00.76  	& 17.82	& 0.90	\\
{\bf \object{Cl* Melotte 22 CFHT-Pl 15}}       &       03 55 12.5      &       +23 17 38.0  	& 18.62	& \nodata	\\
{\bf \object{Cl* Melotte 22 CFHT-Pl 16}}       &       03 44 35.3      &       +25 13 44.0  	& 18.47	& 1.11	\\
{\bf \object{Cl* Melotte 22 CFHT-Pl 17}}      &       03 43 00.2      &       +24 43 52.1  	& 18.47	& 0.96	\\
{\bf \object{Cl* Melotte 22 CFHT-Pl 21}}       &       03 51 25.6      &       +23 45 21.2  	& 18.88	& 1.07	\\
{\bf \object{Cl* Melotte 22 CFHT-Pl 23}}       &       03 52 18.64     &       +24 04 28.41  	& 19.32	& 1.11	\\
{\bf \object{Cl* Melotte 22 CFHT-Pl 24}}       &       03 43 40.29     &       +24 30 11.34  	& 19.38	& 1.12	\\
{\bf \object{Cl* Melotte 22 CFHT-Pl 25}}       &       03 54 05.37     &       +23 33 59.47  	& 19.69	& 1.21	\\
{\bf \object{Cl* Melotte 22 CFHT-Pl-IZ 2141}}   &       03 44 31.29     &       +25 35 14.42  	& 21.88	& 1.14	\\
{\bf \object{Cl* Melotte 22 CFHT-Pl-IZ 2}}     &       03 55 23.07     &       +24 49 05.01  	& 17.81	& 0.90	\\
{\bf \object{Cl* Melotte 22 CFHT-Pl-IZ 23}}    &       03 51 33.48     &       +24 10 14.16  	& 20.30	& 1.10	\\
{\bf \object{Cl* Melotte 22 CFHT-Pl-IZ 26}}    &       03 44 48.66     &       +25 39 17.52  	& 20.85	& 1.20	\\
{\bf \object{Cl* Melotte 22 CFHT-Pl-IZ 28}}    &       03 54 14.03     &       +23 17 51.39  	& 21.01	& 1.23	\\
{\bf \object{Cl* Melotte 22 CFHT-Pl-IZ 4}}     &       03 41 40.92     &       +25 54 23.0  	& 17.82	& 0.96	\\
{\bf \object{Cl* Melotte 22 IPMBD 29}$^{\star}$}         &       03 45 31.3      &       +24 52 48.0  	& 18.35	& \nodata	\\
\object{Cl* Melotte 22 CFHT-Pl-IZ 10}    &       03 51 44.97     &       +23 26 39.47  	& 18.66	& 1.03	\\
\object{Cl* Melotte 22 CFHT-Pl-IZ 1262}  &       03 44 27.27     &       +25 44 41.28  	& 22.47	& 1.23	\\
\object{Cl* Melotte 22 CFHT-Pl-IZ 13}    &       03 55 04.4      &       +26 15 49.3  	& 18.94	& 1.14	\\
\object{Cl* Melotte 22 CFHT-Pl-IZ 14}    &       03 53 32.39     &       +26 07 01.2  	& 18.94	& 1.14	\\
\object{Cl* Melotte 22 CFHT-Pl-IZ 161}   &       03 51 29.43     &       +24 00 36.79  	& 22.32	& 1.35	\\
\object{Cl* Melotte 22 CFHT-Pl-IZ 17}    &       03 51 26.69     &       +23 30 10.65  	& 19.44	& 1.08	\\
\object{Cl* Melotte 22 CFHT-Pl-IZ 19}    &       03 56 16.37     &       +23 54 51.44  	& 19.56	& 1.10	\\
\object{Cl* Melotte 22 CFHT-Pl-IZ 21}    &       03 55 27.66     &       +25 49 40.72  	& 19.80	& 1.17	\\
\object{Cl* Melotte 22 CFHT-Pl-IZ 25}    &       03 52 44.3      &       +24 24 50.04  	& 20.58	& 1.16	\\
\object{Cl* Melotte 22 CFHT-Pl-IZ 29}    &       03 49 45.29     &       +26 50 49.88  	& 21.03	& 1.27	\\
\object{Cl* Melotte 22 CFHT-Pl-IZ 300}   &       03 51 15.6      &       +23 47 05.38  	& 22.1	& 1.18 	\\
\object{Cl* Melotte 22 CFHT-Pl-IZ 31}    &       03 51 47.65     &       +24 39 59.51  	& 21.05	& 1.26	\\
\object{Cl* Melotte 22 CFHT-Pl-IZ 51}    &       03 46 36.24     &       +25 33 36.21  	& 22.59	& 1.24 	\\
\object{Cl* Melotte 22 CFHT-Pl-IZ 7}     &       03 48 12.13     &       +25 54 28.4  	& 18.46	& 1.12	\\
\object{Cl* Melotte 22 IPMBD 25}$^{\star}$         &       03 46 26.1      &       +24 05 10.0  	& 17.82	& \nodata	\\
\enddata
\tablecomments{Observed objects are indicated in bold face, and the $^{\star}$ symbol indicates the binaries.}
\end{deluxetable}

  \subsection{Observations}
Observations were carried out during cycle 12 between July 2003 and August 2004 as part of the HST Snapshot SNAP-9831 program. Each object was observed in the F814W filter, which provides the best compromise between the efficiency, the sensitivity to our cold objects, and the S/N ratio. Only one band was obtained in order to maximize exposure times, minimize the visit times and thus optimize schedulability.

\begin{figure*}
\begin{center}
\includegraphics*[height=0.7\textheight]{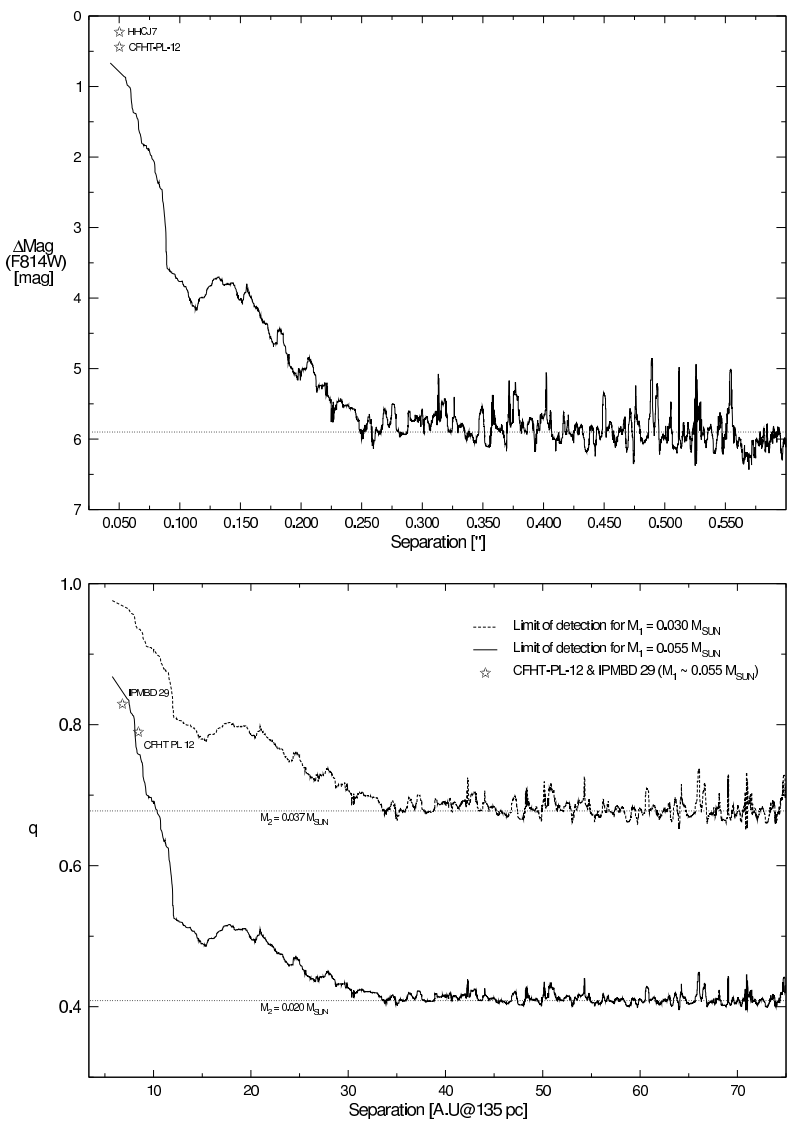}
\caption[Limit of detection with ACS/HRC]{Limit of detection of our ACS/HRC observations.\\{\bf Top panel:} $\Delta$Mag vs angular separation. The curve represent the largest detectable difference of magnitude in the F814W band between the primary and the secondary, as a function of the projected separation. The curve was computed from the average of the 3-$\sigma$ noise measurements in the images. At separation greater than 0\farcs250, we were sensitive to companions 5.9 magnitudes fainter than the primary (dotted line). The two stars indicate the two resolved binaries in this sample.\\ {\bf Bottom Panel:} Same as top panel, but for the mass ratio vs the physical separation. The mass ratios have been computed for 2 different primary masses characteristic of our sample, using the top panel curve and DUSTY models convolved with the HST filters for the mass-luminosity relation. The physical separations have been calculated assuming an average distance of 135~pc. \label{limit_detection_all} }\end{center}
\end{figure*}

Diffraction limited imaging with ACS-HRC at 814~nm gives us a spatial resolution of 0\farcs085. With its 0\farcs027 pixel scale, the ACS-HRC thus provides the required critical sampling of the PSF, which was not the case of the WFPC2/PC camera. Using PSF fitting, we are thus able to resolve even closer companions than in the case of WFPC2. Integration times were 400~s, spread over 4 exposures in CR-SPLIT mode \citep{ACS_INSTR_HANDBOOK}. Figure \ref{limit_detection_all} shows that we are sensitive to companions 5.9~mag fainter than their primary (3-$\sigma$ detection limit), corresponding to a lower limit on the mass ratio between 0.4 and 0.7 at separations greater than 0\farcs250, depending on the brightness of the primary. Considering the total field of view of the ACS camera (26\arcsec$\times$29\arcsec) we were sentitive to companions up to separation as high as $\sim$1700~AU.

\begin{figure*}
\begin{center}
\includegraphics*[width=0.8\textwidth]{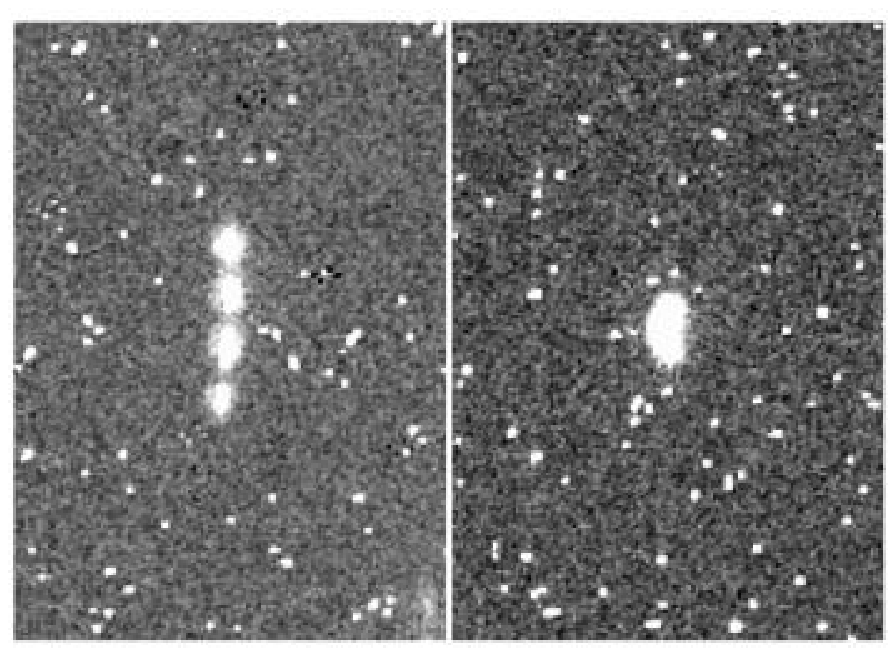}
\caption[Moved ACS exposures]{A problem in the FGS during the acquisition resulted in moved and useless exposures. Left panel: CFHT-Pl-23; Right Panel: CFHT-Pl-24. \label{moved_acs_targets} } 
\end{center}
\end{figure*}

Seventeen objects among the 33 submitted have been observed, but in 2 cases a problem with the guidance sensor resulted in moved exposures, as shown in Figure \ref{moved_acs_targets}. The corresponding images are useless. We thus obtained images for 15 targets, 2 of which were already known binaries.

  \subsection{Data Analysis}

  \subsubsection{Search for the multiple systems}

\begin{figure*}
\begin{center}
\includegraphics*[width=0.8\textwidth]{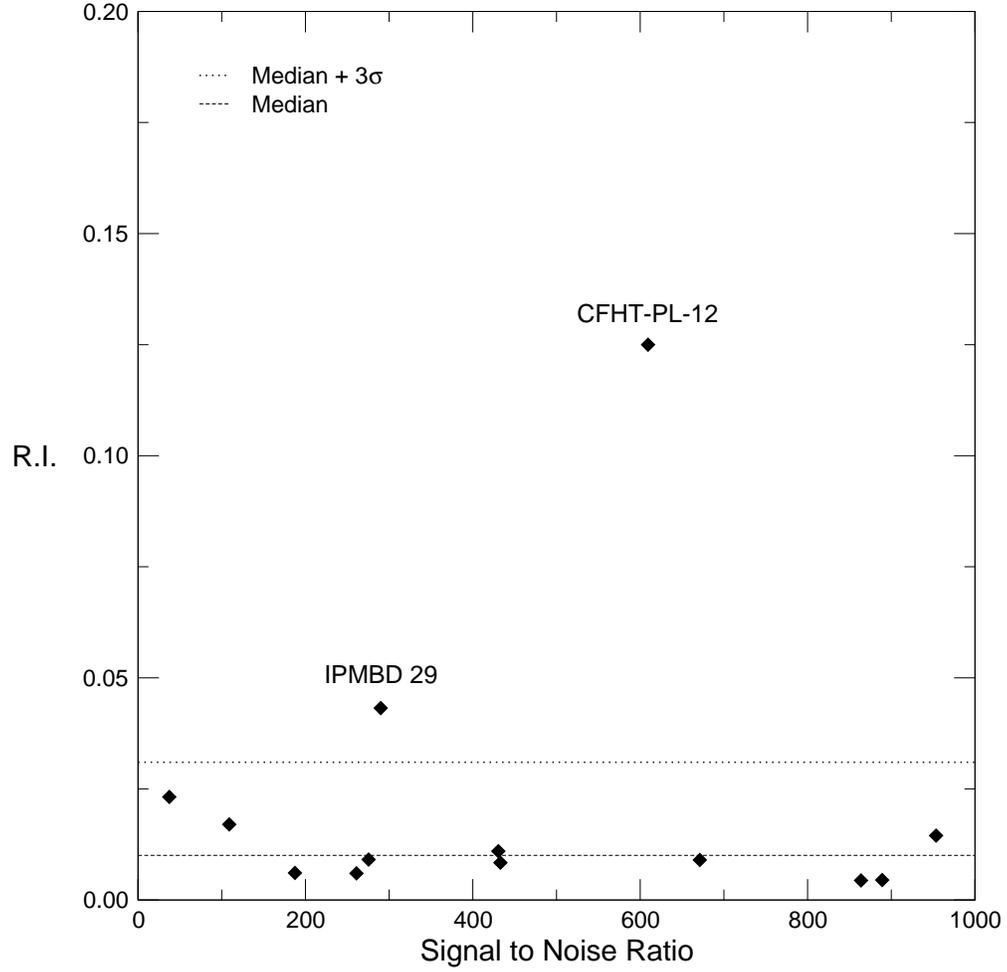}
\caption{Relative intensity of the residuals after PSF subtraction as a function of the SNR. The two binary candidates show clearly higher residuals, above the median+3-$\sigma$ value. \label{ri}}
\end{center}
\end{figure*}

In order to look for multiple systems, we used the same method as described in \citet{2005AJ....129..511B}. Briefly, it consists in a quantitative analysis of the relative intensity of the residuals after PSF subtraction. Any multiple system is expected to show higher residuals than an unresolved one. The technique and its limitations are fully described in the above mentioned article. Figure \ref{ri} shows the result of this analysis. Two systems appear to have clearly higher residuals, indicating that they are very likely to be multiple. These two objects had already been resolved in a previous HST program \citep[see][]{2003ApJ...594..525M}. Some objects at lower SNR also show slightly higher residuals (at about $\sim$1-$\sigma$), but a careful visual inspection of the images and of the PSF subtraction does not show any convincing evidence of multiplicity. As a sanity check, all images have been inspected visually. 

  \subsubsection{PSF fitting}
The ACS-HRC data have been processed with the same PSF fitting program described in \citet{2003AJ....126.1526B}, adapted to ACS-HRC. Briefly: the program performs a dual-PSF fit of the binary, fitting both component at the same time. The relative astrometry and photometry are obtained when the residuals reach their minimum value. The method and its limitations are fully described in \citet{2004PhDT.........5B, 2003AJ....126.1526B}.

\section{Results for the individual objects \label{mutliple_syst}}

We confirm 2 binaries previously discovered in \citet{2003ApJ...594..525M} study, and report no new binary in the angular separation 0\farcs045--0\farcs26  and apparent brightness range 18$<$I$_{C}<$22.8.

Considering the relatively high proper motion of the Pleiades cluster \citep[$\mu_{\alpha}.cos(\delta)=$19.15~mas/yr $\mu_{\delta}=$-45.72~mas/yr; ][]{1999A&A...345..471R}, and the small relative motion of their respective components (see Tables \ref{astrometry_cfhtpl12} and \ref{astrometry_IPMBD_29}), we conclude that CFHT-PL-12AB and IPMBD-29AB are common proper motion pairs. Tables \ref{astrometry_cfhtpl12} and \ref{astrometry_IPMBD_29} show the astrometric measurements of the two objects. For both binaries the separation measured in 2003 is smaller than that measured in 2000. This is an effect of the eccentricity of the orbits and a selection bias due to the resolution limit of the WFPC2 survey.

  \subsection{\index{Cl* Melotte 22 CFHT-Pl 12}Cl* Melotte 22 CFHT-Pl 12}
\begin{figure*}
\begin{center}
\includegraphics*[width=0.8\textwidth]{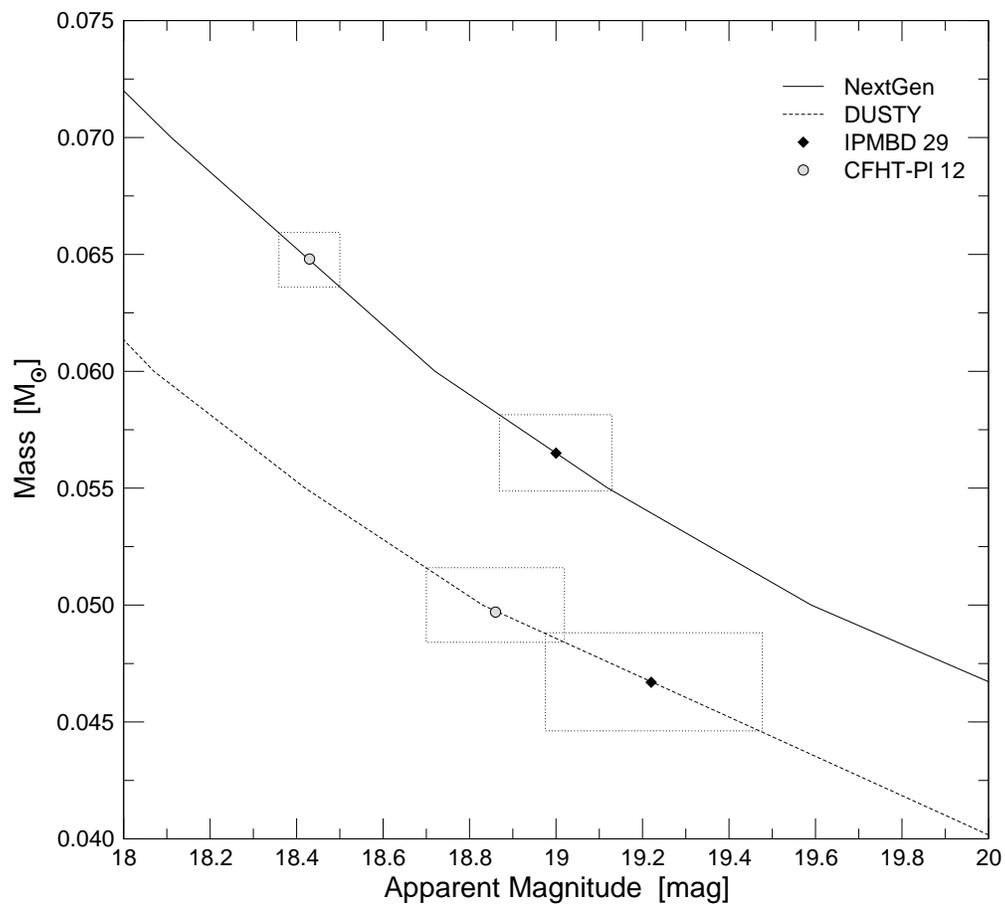}
\caption{Mass vs Apparent Magnitude diagram. The 120~Myr isochrones of the DUSTY and NextGen models are represented together with the measurements we obtained for CFHT-Pl~12 and IPMBD~29, assuming a distance of 135~pc. The propagated uncertainties on the magnitude translate into uncertainties on the mass. These uncertainties are indicated as boxes.\label{cmd}}
\end{center}
\end{figure*}

Cl* Melotte 22 CFHT-Pl 12 is a binary with a separation of 0\farcs062$\pm$0\farcs002 and a position angle (P.A) of 266.7$\pm$1.7\degr\, (14th November 2000), corresponding to a physical separation of 8.4$\pm$0.3~AU at 135~pc. Correcting for a statistical factor of 1.26 as explained in \citet{1992ApJ...396..178F}, it leads to a semi-major axis of 10.5$\pm$0.3~AU. Its proper motion and the presence of Li absorption in its spectrum indicate that it is substellar and belongs to the Pleiades cluster \citep{1998ApJ...499L.199S, 2001A&A...367..211M}. Table \ref{pleiades_bin} gives a summary of its astrometric and photometric properties. Using the NextGen models for the primary and the DUSTY models for the fainter (and therefore cooler) secondary and assuming an age of 120~Myr, we can estimate the masses of each component to be M$_{A}$=0.066$\pm$0.001~M$_{\sun}$ and M$_{B}$=0.052$\pm$0.002~M$_{\sun}$, corresponding to a mass ratio of $q=0.79$ (see Figure. \ref{cmd}). According to Kepler's third laws \citep{1609QB41.K32.......}, the corresponding period is $\sim$99$\pm$5~years. The small relative motion of 15\degr\, in 3 years corresponds to an orbital period of $\sim$70~years, which is of the same order than the orbital period derived from the theoretical masses and the semi-major axis, but a more precise comparison between dynamical masses and theoretical masses requires more astrometric monitoring.

\begin{center}
\begin{deluxetable}{lccccc}
\tablecaption{Relative Astrometry and photometry of Cl* Melotte 22 CFHT-PL 12 \label{astrometry_cfhtpl12}}
\tablewidth{0pt}
\tablehead{
Date           &  Instrument & Sep.       &  P.A                  & $\Delta$Mag  & Filter \\
\textsc{dd/mm/yyyy} &         &   [mas]    &  [\degr]              &              &          \\
}
\startdata
14/11/2000     &   WFPC2     &  62$\pm$3  &  266.7$\pm$4.5        & 0.98$\pm$0.15 & F814W \\
07/11/2003     &   ACS       &  50$\pm$3  &  251.4$\pm$0.75       & 0.43$\pm$0.15 & F814W \\
\enddata
\tablecomments{The difference of magnitude is different at the two epochs. They agree within 2-$\sigma$, but the WFPC2 value should be considered with more caution than the ACS value. The ACS image is indeed much better sampled (the pixel-scale of ACS is twice that of WFPC2). We therefore consider that the ACS value is more accurate.}
\end{deluxetable}
\end{center}

  \subsection{\index{Cl* Melotte 22 IPMBD 29}Cl* Melotte 22 IPMBD 29}
\begin{figure*}
\begin{center}
\includegraphics*[width=0.8\textwidth]{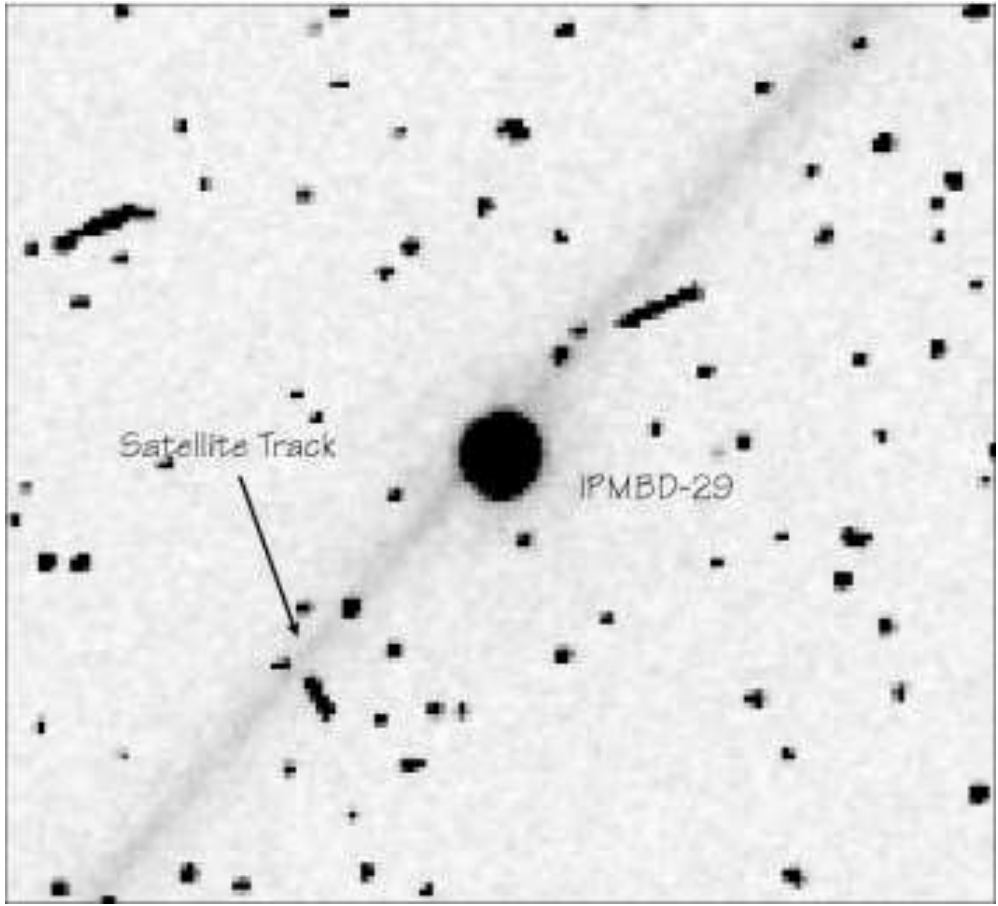}
\caption[Satellite track on the ACS image of Cl* Melotte 22 IPMBD 29]{Satellite track on the ACS image of Cl* Melotte 22 IPMBD 29. Very unfortunately the way of a satellite crossed the field exactly on the position of the target. The corresponding flux is nevertheless relatively small, but might explain part of the $\Delta$Mag difference reported in Table \ref{astrometry_IPMBD_29} \label{IPMBD29_satellite} } 
\end{center}
\end{figure*}

Cl* Melotte 22 IPMBD 29 was confirmed as a Pleiades member via proper motion measurements by \citet{1999MNRAS.303..835H}. It was observed twice: the first time with WFPC2 (18th September 2000), and the second time with ACS (13th December 2003). Table \ref{astrometry_IPMBD_29} gives a summary of the astrometric and photometric properties measured at both epochs. Unfortunately a satellite crossed the field of our ACS image exactly on the target (see Figure \ref{IPMBD29_satellite}). The flux of the satellite track is relatively low. Measuring the number of counts in an area of 11 pixels around the source and in another area centered on the satellite track away from the source, we can estimate that the flux of the satellite track corresponds to less than 5\% of that of the source. The elongation and the duplicity are nevertheless real, since it appears clearly on the 3 individual exposures of the CR-SPLIT that have not been affected by the satellite track. It is moreover confirmed by the previous detection in the WFPC2 image 3 years earlier, with consistent relative astrometry of the two components. 
The difference of magnitude is different at the two epochs. They agree within 3-$\sigma$, but the WFPC2 value should be considered with more caution than the ACS value. The ACS image is indeed much better sampled (the pixel-scale of ACS is twice as good as that of WFPC2), and the separation is below the sampling limit of WFPC2, while it is above that of ACS. We therefore consider that the ACS value is more reliable than the WFPC2 one. Uncertainties on the relative photometry at such short separations should always be considered with caution, since we are much below the diffraction limit of HST at this wavelength. The difference between the measurements obtained with two different instruments on-board HST illustrate the limitations of the PSF fitting.

Cl* Melotte 22 IPMBD 29 is a binary with a separation of 0\farcs050$\pm$0\farcs003 and P.A of 85.6\degr$\pm$0.75\degr\, corresponding to a physical separation of 6.75$\pm$0.4~AU at 135~pc. Correcting for a statistical factor of 1.26 as explained in \citet{1992ApJ...396..178F}, it leads to a semi-major axis of 8.5$\pm$0.5~AU. Using the NextGen models for the primary and the DUSTY models for the fainter secondary and assuming an age of 120~Myr, we can estimate the masses of each component to be M$_{A}$=0.056$\pm$0.002~M$_{\sun}$ and M$_{B}$=0.047$\pm$0.002~M$_{\sun}$, corresponding to a mass ratio of $q=0.83$ (see Figure. \ref{cmd}). According to Kepler's  third laws, the corresponding period is $\sim$77$\pm$9~years. The small relative motion of 5\degr/yr corresponds to an orbital period of $\sim$75~years, consistent with the period derived from the Kepler's laws.

\begin{center}
\begin{deluxetable}{lccccc}
\tablecaption{Relative Astrometry and photometry of Cl* Melotte 22 IPMBD 29 \label{astrometry_IPMBD_29}}
\tablewidth{0pt}
\tablehead{
Date           &  Instrument & Sep.       &  P.A                  & $\Delta$Mag  & Filter \\
\textsc{dd/mm/yyyy} &         &   [mas]    &  [\degr]              &              &          \\
}
\startdata
18/07/2000     &   WFPC2     &  58$\pm$3  &  103$\pm$4.5          & 1.25$\pm$0.15\tablenotemark{a} & F814W \\
13/12/2003     &   ACS       &  50$\pm$3  &  85.6$\pm$0.75        & 0.22$\pm$0.30\tablenotemark{a} & F814W \\
\enddata
\tablenotetext{a}{The difference of magnitude is different at the two epochs. They agree within 3-$\sigma$, but the WFPC2 value should be considered with more caution than the ACS value. We consider the ACS image more reliable than the WFPC2 one.}
\end{deluxetable}
\end{center}

\section{Confirmed photometric binary candidates \label{conf_photom}}

From its position in the H-R diagram, \citet{2003A&A...400..891M} suspected CFHT-Pl-12 to be a brown dwarf binary. Similarly, from their photometric analysis, \citet{2003MNRAS.342.1241P} suspected this object to be multiple. Using our WFPC2 and ACS images, we resolve CFHT-Pl-12 and calculate a mass ratio consistent with the one they derive from the photometry.

\begin{figure*}
\begin{center}
\includegraphics*[width=0.9\textwidth]{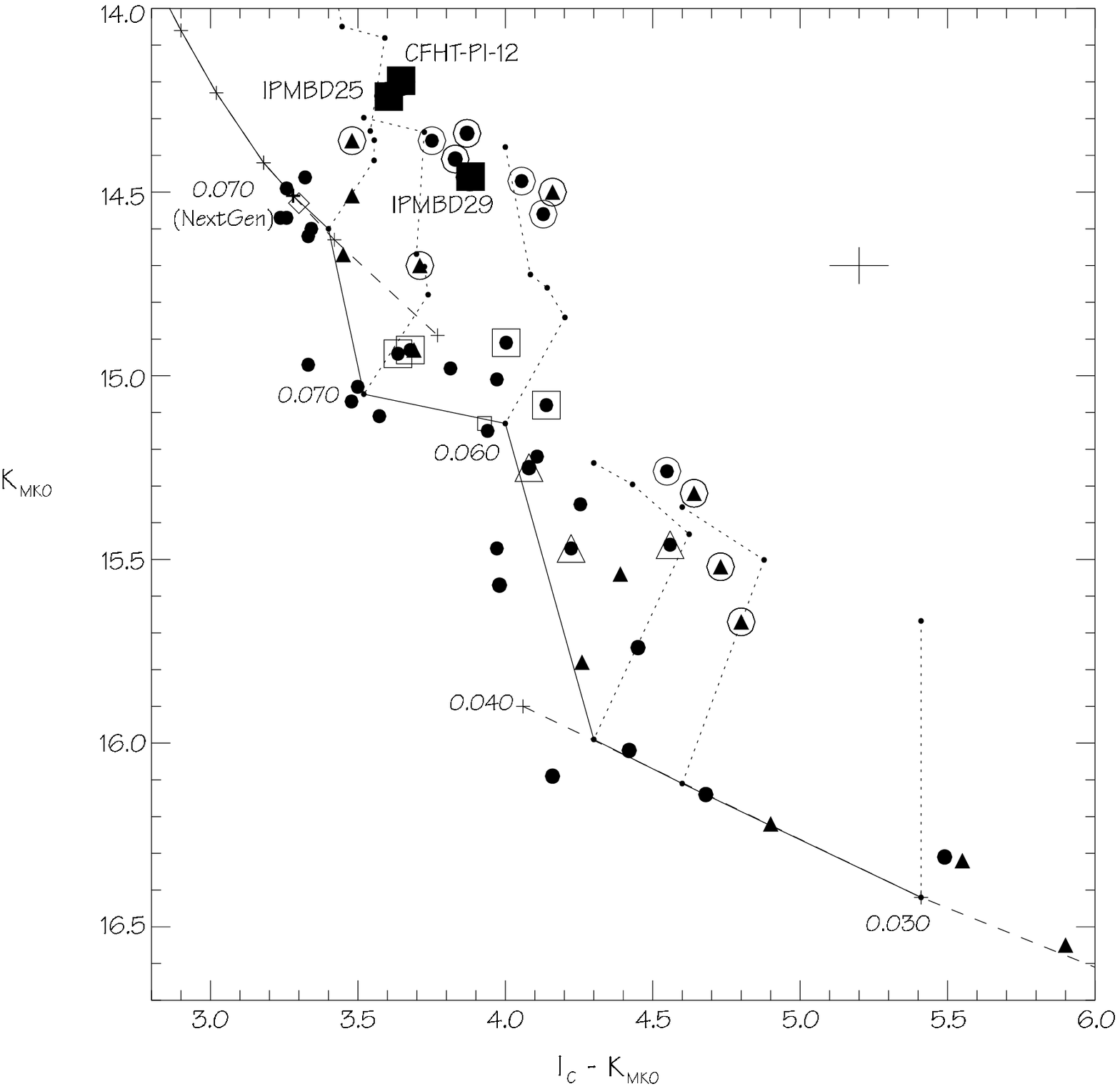}
\caption[$K$ vs $I_{C}-K$ CMD of the Pleiades photometric binary candidates]{$K$ vs $I_{C}-K$ from \citet{2003MNRAS.342.1241P} plus the resolved binaries CFHT-Pl-12, IPMBD-25 and IPMBD-29 \citep[represented by large filled squares; values from][]{1999MNRAS.303..835H}. The symbols mean the same as in \citet{2003MNRAS.342.1241P} paper: circled objects are $IK$ binary candidates, objects overplotted with an open square or triangle are respectively $JK$ or $JHK$ binary candidate. Dashed lines are the NextGen and DUSTY models. Solid and dotted lines are the cluster single and binary star sequences, respectively. Typical uncertainties are indicated. Corresponding masses (in units of solar masses) from the DUSTY models are indicated. The 0.070~M$_{\sun}$ point around K=14.5 is the NextGen model prediction for a 125~Myr isochrone.The two resolved binaries fall on the binary sequence. \label{pinfield} } 
\end{center}
\end{figure*}

It is interesting to note that the two resolved binaries IPMBD-25 and IPMBD-29, which have $I_{C}$ and $K$ photometric measurements available, fall just on the binary sequence of the $K$ vs. ($I_{C}-K$) colour-magnitude diagram (CMD) defined by \citet{2003MNRAS.342.1241P}, as shown in Figure \ref{pinfield}, although they were not included in their study. From this diagram we can predict a mass ratio of 0.6--0.9 for IPMBD-25, very similar to that of CFHT-Pl-12 since the two objects are very close in the diagram, and consistent with the mass ratio we derive from the relative photometry of the two components. Similarly, the CMD predict a mass ratio of 0.7--1.0 for IPMBD-29, in good agreement with the one we derive from the relative photometry of the two components.

\section{Unresolved photometric binary candidates}

From their positions in the H-R diagram, \citet{2003A&A...400..891M} suspected CFHT-Pl-16 to be a brown dwarf binary. It is not resolved in our ACS images. From their photometric study, \citet{2003MNRAS.342.1241P} also classify this object as binary, and derive a mass ratio of about 0.75--1. According to the DUSTY models, this mass ratio corresponds to a difference of magnitude between 0.0$\le\Delta$mag$\le$6~mag in the $I$ band, thus just at/above the limit of sensitivity of our study. This indicates that, if multiple, this system should have a separation less than 5.4--34~AU depending on the flux ratio (see Figure \ref{limit_detection_all} and Table \ref{binary_candidates}).

Due to its peculiar proper motion, \citet{2001A&A...367..211M} suggested that CFHT-Pl-15 might be a multiple system. \citet{2000ApJ...543..299M} found evidence for high residuals after PSF subtraction on their NICMOS image, and suspected the presence of a companion at a separation less than 0\farcs22. Using ACS, we do not resolve any companion at separation larger than 0\farcs040. If multiple, this object should have a separation smaller than 5.4~AU and/or a difference in magnitude larger than 5.9~mag in the F814W band.

From their photometric analysis, \citet{2003MNRAS.342.1241P} suspected CFHT-Pl-25, CFHT-Pl-23 and CFHT-Pl-21 to be binaries. Using our ACS images, we do not find any evidence of companions around these three objects.  \citet{2003MNRAS.342.1241P} also predict mass ratios of $q\sim$1 for CFHT-Pl-23, $q<$0.75--1 for CFHT-Pl-25, and 0.5$<q<$0.7 for CFHT-Pl-21, corresponding to differences of magnitude of respectively 0~mag, $>$0--3~mag, and 3.3--8.8~mag. Together with our ACS study, this constrains the separations of CFHT-Pl-23 to be smaller than 5.4~AU and that of CFHT-Pl-25 to be smaller than $\sim$5.4--13~AU while that of CFHT-Pl-21 should be less than 13~AU (see Figure \ref{limit_detection_all}). Spectroscopic studies would be currently the only way to test the possibility that these objects are binaries. Table \ref{binary_candidates} summarizes this analysis.

\begin{center}
\begin{deluxetable}{lcccc}
\tablecaption{Properties of the unresolved photometric binary candidates \label{binary_candidates}}
\tablewidth{0pt}
\tablehead{
Object           &  $q_{phot.}$   &  I$_{C}$ & $\Delta$mag  &  Limit on Sep.  \\
                 &                &   [mag]  &  [mag]       &  [AU          \\
}
\startdata
CFHT-Pl-16       &  0.75--1.0     &   18.7   &  0.0--6.0   &  $<$5.4--34.0      \\
CFHT-Pl-21       &  0.5--0.7      &   19.0   &  3.5--8.8   &  $<$13.0--34.0  \\ 
CFHT-Pl-23       &  $\sim$1       &   19.3   &  $\sim$0.0  &  $<$5.4          \\
CFHT-Pl-25       &  $<$0.75--1.0  &   19.7   &  $>$0.0--3.5 & $<$5.4--13.0 \\
\enddata
\tablecomments{$q_{phot.}$ is the mass ratio reported by \citet{2003MNRAS.342.1241P} from their photometric study. I$_{C}$ from \citet{2003A&A...400..891M}. $\Delta$mag is obtained using I$_{C}$, $q_{phot.}$, and the DUSTY evolutionary models. The limit on the separation is then derived using Figure \ref{limit_detection_all} }
\end{deluxetable}
\end{center}

\begin{deluxetable}{lccccccccccc}
\tabletypesize{\tiny}
\tablecaption{Results for Pleiades Binary Systems \label{pleiades_bin}}
\tablewidth{0pt}
\tablehead{
\colhead{Name} & \multicolumn{2}{c}{Mag. F814W} & & \multicolumn{2}{c}{Mag. F875LP} & \colhead{Sep.} & \colhead{Sep.} & \colhead{P.A.} & \colhead{M$_{\rm A}$} & \colhead{q} & \colhead{P}\\
\cline{2-3} \cline{5-6}
               & A  & B          &          & A & B                      &   [\arcsec]              &      [AU          &   [\degr]              &  [M$_{\sun}$]                         &   & [yr]                                      }
\startdata
CFHT-Pl 12 & 18.34$\pm$0.11 & 19.32$\pm$0.11 & & 17.57$\pm$0.11 & 18.48$\pm$0.11 & 0.062$\pm$0.002 & 10.5$\pm$0.3  & 266.7$\pm$1.7 & 0.066 &  0.79 &  99 \\
IPMBD 25   & 17.93$\pm$0.09 & 19.38$\pm$0.09 & & 17.22$\pm$0.09 & 18.74$\pm$0.09 & 0.094$\pm$0.003 & 16.0$\pm$0.5 & 340.5$\pm$2.1 & 0.063 &  0.62 & 200 \\
IPMBD 29   & 18.70$\pm$0.15 & 19.95$\pm$0.15 & & 17.81$\pm$0.11 & 19.06$\pm$0.11 & 0.058$\pm$0.004 & 8.6$\pm$0.5 & 103.0$\pm$4.5 & 0.056  &  0.83 &  77   \\
\enddata
\tablecomments{F875LP magnitudes from \citet{2003ApJ...594..525M}. Orbital periods are estimated for circular orbits using Kepler's third law and a distance of 135~pc and are given in years.}
\end{deluxetable}

  \section{Analysis: Binary frequency \label{bin_frequ}}

Our sample of bona-fide brown dwarfs Pleiades members include 15 objects. Two of them were peviously known binaries, and should therefore be excluded from the statistics. This gives an observed visual binary frequency  of $<$7.7\% for separations greater than 5.4~AU and primary masses between 0.030--0.065~M$_{\sun}$. The binary frequency is defined here as the number of binaries divided by the total number of objects in the sample. Upper limit uncertainty is derived as explained in \citet{2003ApJ...586..512B}.

\citet{2003ApJ...594..525M} noticed that the primaries of the only two binaries resolved with WFPC2 are brighter than $I$=18.5~mag, suggesting breaking the statistical analysis in two bins of magnitudes. In the first bin, between 17.7$<I<$ 18.5~mag corresponding to 0.055$<M<$0.065~M$_{\sun}$, they reported a binary frequency of 22$^{+19}_{-8}$\%, with 2 binaries among a sample of 9 objects. In the same magnitude bin, and over the same separation range ($>$7--12~AU, we have 6 new objects and 0 new binary. The combination of the two results gives a total of 2 binaries over 15 objects, leading to a refined binary frequency of 13.3$^{+13.7}_{-4.3}$\%. In the second magnitude bin, between 18.5$<I<$21.0 corresponding to 0.035$<M<$0.055~M$_{\sun}$, \citet{2003ApJ...594..525M} reported 0 binary among a total of 6 objects. In the same magnitude bin and over the same separation range ($>$7--12~AU, we report 5 new objects and 0 new binary. The combination of the two results gives a total of 0 binary over 11 objects, leading to a refined limit on the visual binary frequency of $f_{vis} <$9.1\%. 

In the new separation range that we were able to investigate with ACS, between 5.4--7.0~AU (for the brightest objects only, 17.7$<I<$18.5~mag or 0.055$<M<$0.065~M$_{\sun}$, see Fig. \ref{limit_detection_all}), we report 0 binaries among a total of 6 objects, leading to a limit on the visual binary frequency of $f_{vis} <$16.7\%, consistent with that reported in the separation range between 7--12~AU for the same range of masses.

To summarize, we obtain the following binary frequencies: in the separation range $>$5.4--7.0~AU and in the range of mass between 0.055$<M<$0.065~M$_{\sun}$, we report a visual binary frequency of $f_{vis} = \frac{0}{6}<$16.7\%. In the separation range $>$7-12~AU and in the mass range 0.055$<M<$0.065~M$_{\sun}$, we report a visual binary frequency of $f_{vis} = \frac{2}{15}=$13.3$^{+13.7}_{-4.3}$\%. In the separation range $>$7-12~AU and in the mass range 0.035$<M<$0.055~M$_{\sun}$, we report a visual binary frequency of $f_{vis} = \frac{0}{11}<$9.1\%. Table \ref{pleiades_study} gives an overview of these results.

The three binaries observed in the WFPC2 study all have separations less than 12~AU. The mass ratios are all larger than 0.62.  \index{PPL~15}PPL~15, the spectroscopic binary brown dwarf discovered by \citet{1999AJ....118.2460B}, has a semi-major axis of 0.03~AU and a mass ratio of 0.87. Although this sample is too small for allowing any meaningful statistical study, it is interesting to note that these results are consistent with that obtained in the field for slightly more massive objects, for which a cut-off in the separation range at 20$\sim$30~AU and a possible lack of small mass ratios\footnote{this latter result might be due to observational biases} are observed \citep[$q\le$0.5][]{2005ApJ...621.1023S,2003AJ....126.1526B, 2003ApJ...587..407C,2003AJ....125.3302G}.

\renewcommand{\arraystretch}{1.5}
\begin{deluxetable}{lcccccc}
\tabletypesize{\small}
\tablecaption{Visual Binary Frequency measured in successive studies. \label{pleiades_study}}
\tablewidth{0pt}
\tablehead{
\colhead{Ref.}             & N$_{Objects}$ & N$_{Binaries}$ & Sep. Range  & Mass range\tablenotemark{a}   & Sensitivity & Bin. Freq.\tablenotemark{b} \\
                           &               &                &  [AU      & [M$_{\sun}$] &  ($q_{min}$)\tablenotemark{c} &  \\         
} 
\startdata
\citet{2000ApJ...543..299M} & 34           &    0           & $>$24   &  $>$0.090        &  0.6                           & $<$3\% \\
\hline
\citet{2003ApJ...594..525M} & 13           &    2           & $>$7--12   &  0.040--0.065    &  0.45--0.9                     & 15$^{+15}_{-5}$\% \\
\hline
\citet{2003ApJ...594..525M} &  9           &    2           & $>$7--12   &  0.055--0.065    &  0.45--0.9                     & 22$^{+19}_{-8}$\% \\
{\bf ACS+\citet{2003ApJ...594..525M}} & 15 &    2        & $>$7--12   &  0.055--0.065    &  0.45--0.9                     & 13.3$^{+13.7}_{-4.3}$\% \\
{\bf this ACS study}              & 6            &    0           & $>$5.4--7.0  &  0.055--0.065    &  0.9                           & $<$16.7\% \\
\hline
\citet{2003ApJ...594..525M} &  6           &    0           & $>$7--12   &  0.035--0.055    &  0.45--0.9                     & $<$16.7\% \\
this ACS study              &  5           &    0           & $>$7--12   &  0.035--0.055    &  0.45--0.9                     & $<$20.0\% \\
{\bf ACS+\citet{2003ApJ...594..525M}} & 11 &    0        & $>$7--12   &  0.035--0.055    &  0.45--0.9                    &  $<$9.1\% \\

\enddata
\tablenotetext{a}{ for the primary}
\tablenotetext{b}{ Binary frequency defined as N$_{binaries}/$N$_{Objects}$}
\tablenotetext{c}{ Range of sensitivity to lower mass companions, expressed as the minimum mass ratio $q=M_{2}/M_{1}$ to which the observations were sensitive.}
\end{deluxetable}

 \section{Discussion \label{discussion}}

  \subsection{Properties of multiplicity and the mass}
Both the present ACS study and \citet{2003ApJ...594..525M} WFPC2 study suggest that there might be an important change in the properties of multiplicity within the brown dwarf regime. Although statistically inconclusive because of the small number statistics and the relatively large uncertainties, the binary fractions in the two ranges of mass 0.035--0.055~M$_{\sun}$ ($f_{vis}<$9.1\%) and 0.055--0.065~M$_{\sun}$ ($f_{vis}=$13.3$^{+13.7}_{-4.3}$\%) seems to be notably different. This could mean that the brown dwarf binaries at lower masses are tighter, as already suggested by \citet{2003ApJ...587..407C}, and therefore were not resolved by any of the ACS or WFPC2 studies. The small separations reported for the 3 field binary T-dwarfs currently known \citep[all in the range 0--2.7~AU][]{2003ApJ...586..512B,2004A&A...413.1029M} are consistent with this result. 

  \subsection{Properties of multiplicity and the environment \label{binarity_environment}}

Figure \ref{bin_freq_vs_spt_Pleiades+field} shows that the observed binary frequency among the Pleiades brown dwarfs (13.3$^{+13.7}_{-4.3}$\% for separation greater than 7--12~AU is similar to the values reported in the field: 1) for slightly more massive objects \citep[see][10$\sim$15\% of late-M, L-dwarfs]{2005ApJ...621.1023S,2003AJ....126.1526B,2003ApJ...587..407C,2003AJ....125.3302G}; 2) for field brown dwarfs, as reported by \citet[][9$^{+15}_{-4}$\% for T5 to T8 field brown dwarfs]{2003ApJ...586..512B}.

This indicates that the statistical properties, and therefore the formation and evolution processes, of field and Pleiades binary brown dwarfs are probably similar. This would imply that the evolution processes of very low mass binaries do not depend much on the age after 120~Myrs, as expected. The formation, the evolution and, possibly, the disruption of binaries responsible for the low rate of binaries and the cut-off in the separation range would thus have to occur during the early stages of the cluster, when its density and the probability of gravitational encounters are higher. N-body simulations performed by \citet{1995MNRAS.277.1491K,1995MNRAS.277.1522K} have shown that in dense stellar clusters, such as the Pleiades during its early stages, the binary fraction could drop from 100\% to $\sim$50\% in less than 1~Myr. More recent hydrodynamical simulations undertaken by \citet{2005MmSAI..76..223D} led to similar conclusions, with a typical decay-time for multiple systems of $\sim$10~Myr, consistent with the preliminary conclusion we draw here.

In their numerical simulations of the dynamical interactions in stellar clusters, \citet{2003A&A...400.1031S} show that the different properties cited above (binary fraction and distribution of separation) can be nicely reproduced when considering a small-N cluster model (N$<$10) where stars and brown dwarfs form from progenitor clumps. Choosing specific clump and stellar mass spectra, they were able to generate a cluster with an IMF consistent with that observed. Using Monte-Carlo simulations they could then study the small-N cluster decay dynamics and compute the properties of brown dwarfs and brown dwarf binaries. Their study shows that a simple gravitational point-mass dynamics, with weighting factors for the pairing probabilities as a function of the mass evaluated in the first of a two step process, gives results consistent with the observations over the entire range of mass. In particular, they obtain a binary fraction for brown dwarfs of 8--18\%, consistent with the binary fraction we report here (13.3$^{+13.7}_{-4.3}$\%). They also model a distribution of separation in remarkable agreement with that reported for the field brown dwarfs and for the three Pleiades binaries of our study, with a peak around 4~AU and most ($\sim$85\%) objects with separations less than 20~AU. On the other hand, they produce a flat distribution of mass ratio in the range 0.2$<q<$1.0, which is apparently not observed in the field and in the Pleiades. \citet{2005ApJ...621.1023S, 2003AJ....126.1526B, 2003ApJ...586..512B,2003ApJ...587..407C,2003AJ....125.3302G} showed that their observations in the field, although statistically incomplete, suggest that there is a preference for equal mass systems. \citet{2003A&A...397..159H} showed also that the mass ratio distribution of spectroscopic binaries among field and Pleiades F--G dwarfs is not flat but bimodal. Finally, in a similar recent study performed on the decay of accreting\footnote{\citet{2003A&A...400.1031S} simulations were purely dynamical, neglecting accretion, but considering small-N clusters rather than triple systems} triple systems, \citep{2005ApJ...623..940U} shows that they are also able to reproduce nicely both the distribution of separation observed for field brown dwarfs, with a cut-off around 20~AU.

  \subsection{Photometric binary frequency \label{photom_freq}}
Our work allows the measurement of the binary frequency among brown dwarfs in the Pleiades Open Cluster for separations greater than 7~AU masses between 0.055--0.065~M$_{\sun}$, and mass ratios in the range 0.45--0.9$<q<$1, with $f_{vb}=$13.3$^{+13.7}_{-4.3}$\% (visual binaries). We will compare this result to that obtained for slightly more massive objects by \citet{2003MNRAS.342.1241P} via the study of binary sequences in colour-magnitude diagrams.

The results of \citet{2003MNRAS.342.1241P} do not agree with the observations we report here. From their study of $IK$, $JK$ and $JHK$ colour-magnitude diagrams, they measure a binary frequency of 50$^{+11}_{-10}$\% for brown dwarfs in the Pleiades in the mass range 0.05--0.07~M$_{\sun}$ with mass ratio between 0.5$<q<$1.0, thus comparable to the ranges covered by our study. This result is much higher than any of the two values reported in our WFPC2 and ACS studies. If correct, these results together would imply that most ($\sim$85\%) of the Pleiades brown dwarf binaries in the range 0.055--0.065~M$_{\sun}$ and 0.5$<q<$1.0 have separations less than 7~AU. From their simulations, \citet{2005MNRAS.tmpL..67M} have recently shown that the spectroscopic binary fraction might be as high as 17--30\% for separations less than 2.6~AU This value, together with the one we report for separations greater than 7~AU adds up to 30--43\% for objects with separations less than 2.6~AU or greater than 7~AU (with a gap between the two). Over the whole separation range, it probably adds up to a binary fraction close to that reported by \citet{2003MNRAS.342.1241P}. On the other hand, a recent spectroscopic surveys among Cha~I brown dwarfs \citep[][no binary candidate out of a sample of 10 objects]{2005astro.ph..9134J} show that the spectroscopic binary fraction seems to be relatively low at young ages.

If confirmed by spectroscopic surveys, it would contrast with the results obtained for late type G--K dwarfs in the Pleiades and for early-M dwarfs in the field. \citet{1992A&A...265..513M} found indeed that only $\sim$30\% of the G--K Pleiades binaries have separations smaller than 5~AU. Similarly, \citet{2004ASPC..318..166D,2003sf2a.confE.248M} found that only $\sim$30\% of the early-M field binaries have separations smaller than 5~AU. These two values are much smaller than the above mentioned 85\%. Assuming that the properties of brown dwarf binaries in that range of masses are similar to that of field or Pleiades late type stars is of course a strong assumption, although we showed in Section \ref{binarity_environment} that the current results tend to confirm it.

The discrepancy between the photometric binary frequency and our visual binary frequency cannot be due to the companions we missed because of their small mass ratios, since the study of \citet{2003MNRAS.342.1241P} is sensitive to a similar range of mass ratio as our study. Moreover \citet{2003A&A...397..159H} found that $\sim$60\% of the F--G Pleiades spectroscopic binaries have a mass ratio larger than 0.5, and \citet{2004ASPC..318..166D,2003sf2a.confE.248M} report that $\sim$75\% of the field early M-dwarfs have a mass ratio larger than 0.5. If once again we make the assumption that field and Pleiades late type binaries have similar properties to Pleiades brown dwarfs binaries, we should have missed between 25--40\% of the multiple systems ``only'', leading to a corrected binary fraction of 15--19\%, still far from the 50\% reported by \citet{2003MNRAS.342.1241P}. 

In addition to the spectroscopic binaries we miss, we suspect that the large discrepancy between the observations we report and the photometric binary frequency of \citet{2003MNRAS.342.1241P} could be due to a combination of the following effects:
\begin{itemize}
\item[-] underestimations of the photometric uncertainties, and of possible intrinsic photometric variability due, for example, to weather effects or magnetically driven surface features. Weather effects are known to be producing variability in the luminosity, up to 0.05~mag in I as observed by \citet{2001A&A...367..218B, 2001ApJ...557..822M}, and magnetically driven surface features modulation of up to 0.1~mag in J \citep[for young Cha-1 brown dwarfs,][]{2003ApJ...594..971J}. 
\item[-] spread in the age of the objects. According to the DUSTY evolutionnary models, a spread in the age between 80 and 125~Myr translates into differences of magnitude of up to 0.1~mag in I.
\item[-] contamination by field objects. Only 14 of 39 brown dwarfs of their sample have been confirmed as cluster members by proper motion and/or Li detection, while all the objects of our sample have been confirmed by one or both tests. The remaining 25 objects (64\% of the sample) have been classified as brown dwarfs on the only basis of their photometric properties. From their photometric (I vs I-Z) and proper motion surveys, \citet{2003A&A...400..891M,2001A&A...367..211M} estimated that the contamination by foreground M-dwarfs in their sample of Pleiades brown dwarfs can be as high as 30\%. From a three colour photometric study (I,Z, and K), they estimate the remaining contamination to be of the order of 10\%. A similar non-negligible level of contamination could be expected in \citet{2003MNRAS.342.1241P} sample and explain some of the red objects identified as binaries. Since the contaminating objects would be foreground (i.e closer) M-dwarfs, most of them would indeed appear close to the Pleiades binary sequence. The binary \object{CFHT-Pl-18} is an example of such contaminating objects \citep{2000ApJ...543..299M}.
\item[-] effect of rotation: brown dwarfs are known to be fast rotators \citep{2004A&A...419..703B}, and a correlation between the rotation and the luminosity, by up to 0.1~mag, could affect the colours of some objects, as measured by \citet{1982Msngr..28...15V}. Deformation of the objects due to their fast rotation can produce variable light curves. A rapidly rotating brown dwarf seen pole-on may be reddened enough to perhaps be identified as a binary by the photometric technique.
\item[-] contamination by non-physical pairs in unresolved blends
\end{itemize}

\begin{figure*}
\begin{center}
\includegraphics[width=\textwidth]{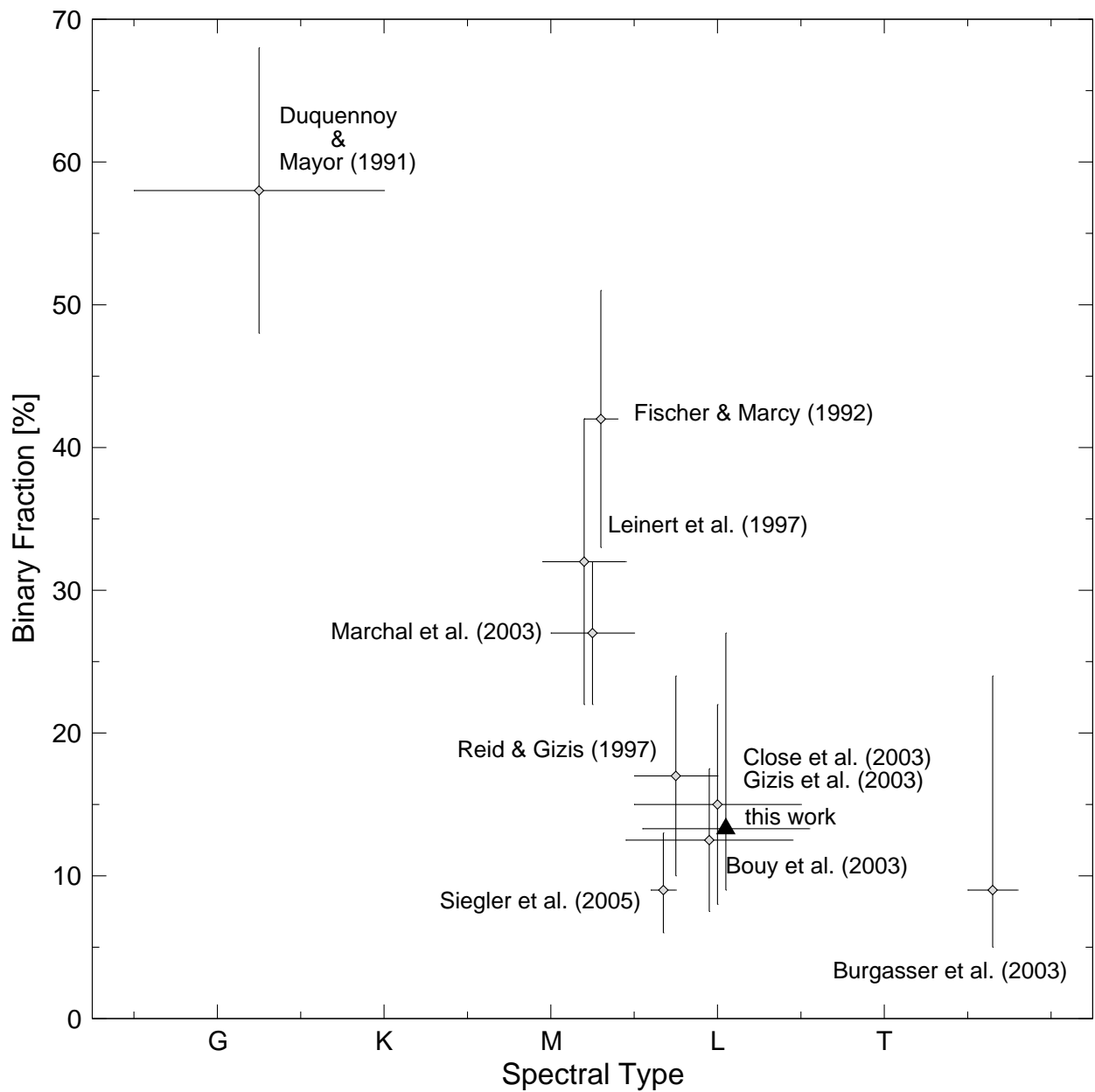}
\caption[Binary Frequency vs Spectral type in the Pleiades: comparison with the field]{Binary frequency as function of the spectral type in the field and in the Pleiades. The value reported in the present work is indicated with a black triangle, while other results for field objects are represented with grey diamonds. The values for spectral types later than M5 are upper limits and do not cover the same ranges of mass ratio and separation than the studies for earlier spectral types, and a direct comparison between the two is not correct. Some points have been slightly shifted ($\pm$0.5 spectral class) to make the figure more clear. \label{bin_freq_vs_spt_Pleiades+field} }
\end{center}
\end{figure*}

The binary frequency we report here for brown dwarfs in the Pleiades is consistent with that observed for similar objects, similar separation and mass ratio ranges than in the field, as shown in Figure \ref{bin_freq_vs_spt_Pleiades+field}. It is comparable to that of slightly more massive field late-M/early-L dwarfs, and close to the frequency observed for field T-dwarfs, which have masses comparable to the brown dwarfs of our Pleiades sample.

Deep spectroscopic surveys on unbiased samples should provide answers to these questions and determine how many small mass ratio/small separation binaries we missed.

  \subsection{Separations and mass ratios}

In his statistical analysis of the photometric binary properties in the Pleiades, \citet{1999A&A...346...67K} shows that the distribution of mass ratios for late type stars should be similar to that in the field. The distribution is expected to be bimodal, with a major peak at $q$=0.4 and a minor one at $\sim$1. In a more recent observational study of unbiased samples of spectroscopic binaries of F to K dwarfs in the field and in the Pleiades cluster, \citet{2003A&A...397..159H} refine the results of \citet{1999A&A...346...67K} in the range of periods shorter than 10~yrs. They report a mass ratio distribution with a primary peak at $q$=1, decreasing towards smaller mass ratios, with a broad secondary peak around $q$=0.4. They observe no difference between the distributions of mass ratio of F--G and K stars, and find that these are identical in the field and in the Pleiades. 

If confirmed, the lack of multiple systems with small mass ratios would then imply a major difference between the distributions of mass ratios (and therefore the formation and evolution processes) of late type stars and brown dwarfs. The current studies are inconclusive regarding that question since the observed lack might well be due to a combination of the following reasons:
\begin{itemize}
\item[-] the bias toward bright magnitudes in favor of binaries with large mass ratios \citep{1924Oepik}
\item[-] the current limit of sensitivity: $q>$0.4 for separation larger than 30~AU and only $q>$0.7 for separations larger than 10~AU (see Figure \ref{limit_detection_all})
\end{itemize}

Deep spectroscopic surveys on unbiased samples should allow to answer these questions, and see how many binaries of small mass ratios and small separations we missed.

\section{Conclusions}
Our new high angular resolution survey for brown dwarf binaries leads to a visual binary fraction in the Pleiades of 13.3$^{+13.7}_{-4.3}$\% for separations larger than 7~AU mass ratio between 0.45--0.9, and masses between 0.055--0.65~M$_{\sun}$. The preliminary results show that there might be a difference in the properties of multiplicity within the brown dwarf regime itself, with smaller separations at smaller masses. The binary frequency we report here is a lower limit of the overall binary frequency. It is much lower than the value reported by \citet{2003MNRAS.342.1241P} for photometric binaries over a slightly higher range of masses in the Pleiades, but a similar range of mass ratio. As suggested by the recent results of \citet{2005MNRAS.tmpL..67M}, the difference could well be due to the spectroscopic binaries missed in our survey. While several surveys looking for visual binaries have already been successfully performed, spectroscopic surveys are only starting to provide results. \citet{2005MNRAS.tmpL..67M} results, as well as the present study, show that there is strong need for such systematic surveys looking for close companions, in the Pleiades but also in the field or in star forming regions. The large difference between the results of the two above mentioned independent and complementary studies, and the remaining uncertainties on the overall binary frequency must remind us that any value of the multiplicity fraction must be very carefully used, and {\bf always considered within its limits (separation range, mass ratio range, mass range)} before a meaningful comparison with other binary frequencies or theoretical predictions can be done.

\acknowledgments

We are grateful to the STSci team and  in particular to our program coordinator Tricia Royle for their kind and efficient support. We also thank our anonymous referee for his comments and corrections. This research is based on observations with the NASA/ESA Hubble Space Telescope, obtained at the Space Telescope Science Institute, and was funded by HST Grants SNAP-9831. 



Facilities: \facility{HST(ACS)}





\end{document}